\documentclass[11pt,A4]{article}
\usepackage[utf8]{inputenc}
\usepackage[english]{babel}
\usepackage{natbib}
\usepackage{graphicx}
\usepackage{amssymb}
\usepackage{paralist}
\usepackage{aas_macros}

\usepackage{wrapfig}
\usepackage{caption}
\usepackage{xcolor}

\title{PHEMTO\\
Polarimetric High Energy Modular Telescope Observatory}
\author{Laurent P. et al.}
\date{August 2019}

\begin{document}

\maketitle
\vspace{5cm} 
\begin{center} 
\Large
\textbf{White Paper submitted to ESA Voyage 2050 Call} \linebreak

\end{center} 

\vspace{5cm} 
\normalsize
\noindent \textbf{Contact Person}: \\
\newline

\noindent Name:	Philippe Laurent \\
Dept.: CEA/DRF/IRFU/DAp, CEA Saclay, France \\
Email: philippe.laurent@cea.fr \\
Phone: +33-1-69086140 \\

\newpage

\section{Introduction}
With the opening of the X and gamma--ray windows in the sixties, thanks to to sounding rockets and satellite-borne instruments, extremely energetic and violent phenomena were discovered and subsequently found to be ubiquitous in the Universe. The existence of compact objects, in particular black holes, was unveiled. In addition to their obvious interest as objects predicted by the General relativity, it has also been realized that black holes of all masses have a fundamental role in shaping the observable properties of the Universe. They are also now central in questions about the evolution of the Universe.

Much more recently, the advent of neutrino and gravitational waves astronomy has led to fundamentally new observations of highly time-variable phenomena in the high-energy domain. A particular example is the hard X-ray INTEGRAL/SPI/ACS and Fermi/GBM detection of the binary neutron star merger GW170817 discovered by gravitational wave interferometers, which has shown that multi-messenger astronomy brings new insights into these violent phenomena.

Observations in the high energy domain are fundamental for understanding how matter is organized and behaves around black holes; unravelling how these extreme objects influence their environments on a very large scale; and finding the still elusive obscured massive objects in the center of galaxies. Other major problems in contemporary astrophysics, such as the understanding of acceleration processes at shocks of all sizes (those of pulsar wind nebulae, supernova remnants, but also at larger scales those of Active Galactic Nuclei radio lobes) in relation to the origin of cosmic-rays, or the definitive characterization of the debated non-thermal X-ray energy content of clusters of galaxies, also requires observations at very high energies.

An observatory type medium mission operating from around 1 keV to about 600 keV can provide direct insights into these major questions. The essential characteristics will be a coverage of the full energy range by telescopes featuring a large throughput and high angular resolution optics, coupled to a compact focal plane assembly, with excellent imaging resolution and spectroscopy. In addition, the mission will provide unique polarimetry measurements in the hard X-ray domain, an important new diagnostic tool at energies for which the non-thermal processes dominate. Acceleration phenomena, non thermal processes and magnetic field topology will be probed with the new and independent observables that are the polarization degree and the position angle.

The Polarimetric High-Energy Modular Telescope Observatory (PHEMTO) is designed to have performance several orders of magnitude better than the present hard X-ray instruments. This, together with its angular resolution around one arcsecond, and its sensitive polarimetry measurement capability, gives to PHEMTO the improvements in scientific performance needed for a mission in the 2050 era.

The PHEMTO observatory will be based upon two focusing devices, an soft/hard X-ray mirror at lower energy, in the continuity of those developed for the ESA ATHENA, NASA NuSTAR and JAXA Hitomi missions and a Laue lens at higher energy, similar to those designed for the ASTENA mission, focusing X–rays onto two identical focal plane detector systems. The gain in the maximum energy that can be focused is also achieved by having a up to 100 m focal length obtained in a formation flying configuration.  


The detailed science objectives are given in the next section, which ends with a table of the scientific requirements. This is followed by more technical sections, which describe a possible model payload and its implementation. 

\newpage

\section{Scientific Objectives}

The PHEMTO scientific objectives are described in detail below. These objectives define the scientific requirements presented in Table~\ref{table_req} of Section~3. 
\vspace{0.3 cm}

\subsection{Black Hole Census, Cosmic X-Ray Background and obscured accreting AGN}

\begin{figure}
\begin{minipage}[t]{14.5cm}
 \begin{minipage}[t]{9cm}
  \begin{flushleft}
  \includegraphics[width=9cm]{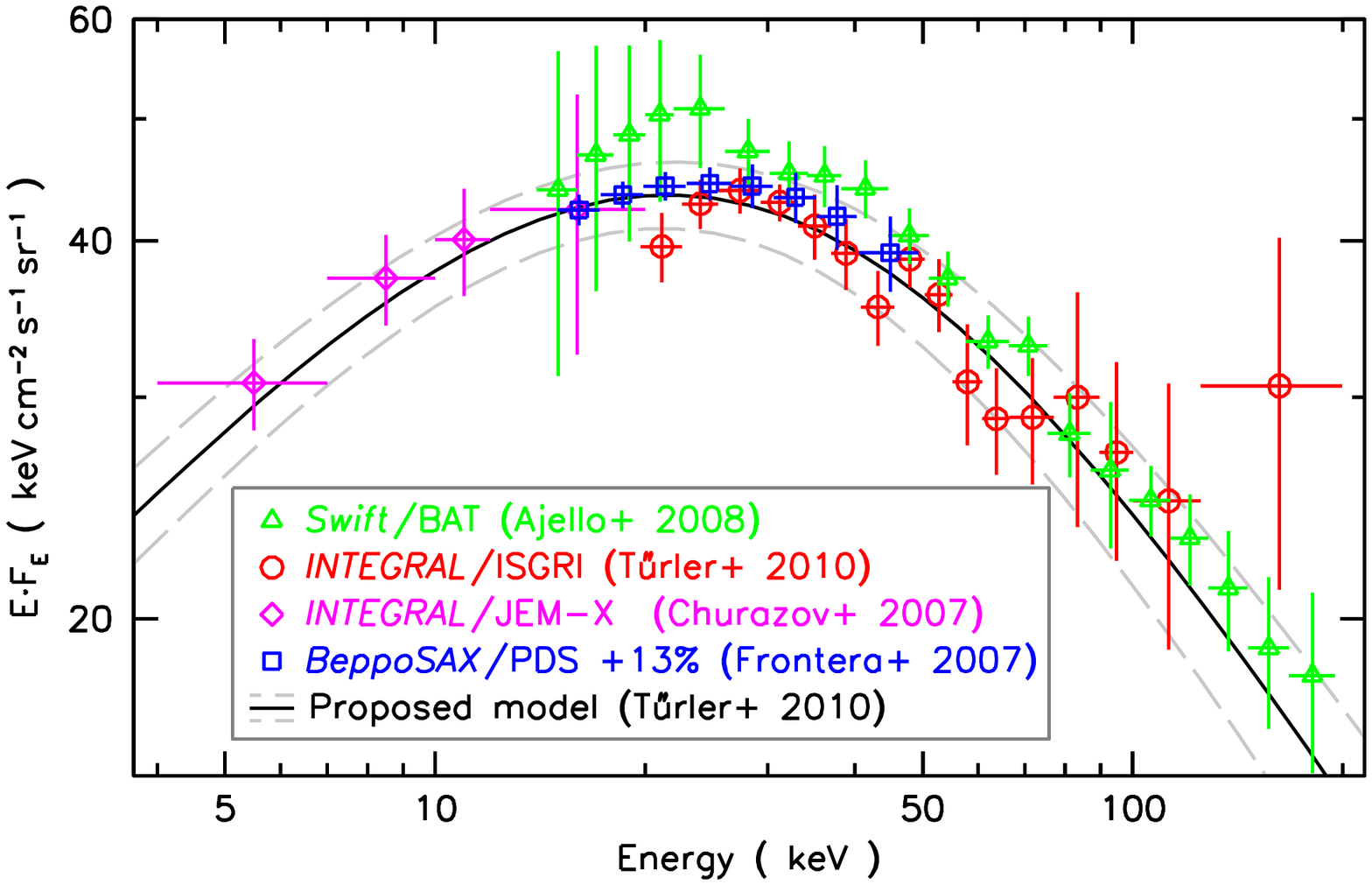}
  \end{flushleft}
 \end{minipage}
 \hfill
 \begin{minipage}[t]{6cm}
  \begin{flushright}
  \includegraphics[width=6cm]{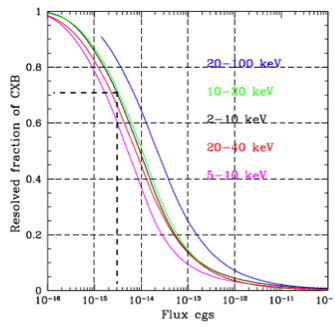}
  \end{flushright}
 \end{minipage}
\end{minipage}
\caption{The left panel shows measurements of the cosmic X-ray background by BeppoSAX, Swift,
and INTEGRAL. The continuous line shows the best fitting model derived by \citet{2010A&A...512A..49T}. Figure courtesy of Marc T\"urler. The right panel indicates the resolved fraction of the CXB as a function of flux limit in different energy bands. At the confusion limit, PHEMTO will resolve 70 \% of the CXB at its peak around 30~keV.} \label{fig:CXB}
\end{figure}

Active Galactic Nuclei (AGN) play a key role in the course of galaxy formation and evolution through interaction with their host galaxies \citep{Beckmann2012}. The Cosmic X-ray Background (CXB) is the fossil of the emission by accretion in the Universe. Its spectrum (Fig.~\ref{fig:CXB} left) tells us that most accretion in the Universe takes place in obscured environments \citep{1999MNRAS.303L..34F} and matching its energy density to the local Super Massive Black Holes (SMBH) mass density reveals that current surveys miss about half of the existing AGN (probably the most obscured ones). Hence, to understand the cosmic history of accretion and its influence on galaxy formation it is mandatory to find and characterize the population of obscured and heavily obscured AGN.
However, this population is so far only loosely constrained. If the hydrogen column density ($N_{\rm H}$) does not exceed a few times $10^{24} \rm \, cm^{-2}$ the nuclear radiation is still visible above 10~keV and the source is called “mildly” Compton thick (CT). If $N_{\rm H} > 10^{25} \, \rm cm^{-2}$ (“heavily” CT) the entire spectrum is depressed by Compton recoil. In both cases, a strong iron line (Equivalent Width = 1–2 keV) and a Compton reflection continuum, peaking around 30~keV, are almost invariably observed. Present X-ray surveys miss most of the highly obscured, but still strongly accreting objects, even deep surveys below 10~keV are inefficient to detect them \citep[e.g.][]{2006A&A...451..457T}. NuSTAR has recently been able to account for $\sim 35\%$ of the X-ray sources responsible for the CXB peak emission, but only through 2-band photometry in hard X-rays and with large uncertainties \citep{2016ApJ...831..185H,2018ApJ...867..162M}. The handful of objects so far discovered in hard X-rays and subsequently identified by e.g. Swift, INTEGRAL, NuSTAR and Chandra \citep{2009A&A...505..417B,2013MNRAS.433.1687M,2014MNRAS.443.1999B}, may represent just the tip of the iceberg as they belong to the very local Universe, while highly obscured objects may well be common at high redshift \citep[e.g.][]{2007A&A...463...79G,2016ApJ...831..185H}.
The PHEMTO main contribution in this field will be the discovery and the characterization of the sources making the strongest contribution to the peak of the CXB around 30~keV. Some of these sources may be already present in deep Chandra and XMM–Newton surveys, as well as in deep mid-infrared surveys, however only sensitive observations at 30~keV can 1) unequivocally identify them as hard X-ray sources and strong contributors to the CXB; 2) quantify their volume density as a function of the Cosmic time; 3) constrain the physics and the geometry of the obscuring matter.

The right panel of Fig.~\ref{fig:CXB} shows the resolved fraction of the CXB as a function of flux. Taken at face value, the plot implies that at the 200~ks PHEMTO flux limit 70\% of the CXB is resolved in sources (compared to a resolved fraction of $\sim 5\%$ in current Swift and INTEGRAL surveys). This fraction, easily achievable with PHEMTO, is comparable or even larger than that already resolved by XMM and Chandra between 5 and 10 keV, and still larger by a factor of $\sim 2$ compared to NuSTAR. In a deep 1 Ms exposure, our simulations show that PHEMTO would be able to measure the X-ray spectral characteristics of the faintest detectable sources up to $z \simeq 3$ and down to $L(2–10 \, \rm keV)$ less than a few $10^{44} \rm \, erg \, s^{-1}$, telling apart unabsorbed, Compton thin and CT objects with high significance, and recovering the intrinsic absorption with less than 20\% uncertainty for Compton thin objects.

\subsubsection{Compton thick QSOs up to z $\approx 1$}

Moderately obscured ($N_{\rm H} \leq 10^{24}\, \rm cm^{-2}$), high-luminosity QSOs have been discovered by previous X-ray satellites. Current X-ray spectra are rather poor, and the uncertainties on the column densities are large. Some of these objects may even be CT but current data cannot tell for sure. PHEMTO spectra will easily distinguish between Compton thin and Compton thick AGN. For a moderately bright ($L_X = 10^{44} \rm \, erg \, s^{-1}$) Compton thick AGN at redshift $z = 1$, the absorbing column density can be determined with an error of 25 \% within a 100~ks observation. Thus PHEMTO can and will for the first time detect and measure Compton thick AGN up to cosmologically relevant redshifts in larger numbers, both through serendipitous surveys and targeting candidates from XMM-Newton, Chandra, and NuSTAR Surveys.

\subsubsection{Deep surveys}
To resolve the CXB at the energy where its contribution reaches its maximum is one of the main goals of the PHEMTO mission. PHEMTO holds the key to uncover and detect directly most SMBH accretion luminosity in the Universe. 
PHEMTO will allow the detection of more than 25~sources per field in the 10–40~keV band in 200 ks of observation time ($\sim 15$ for 100~ks). About a quarter of these sources should be highly obscured AGNs. Most of them will be Seyfert~2 galaxies, but one or two sources per field should be high luminosity type~2 QSOs \citep{2018MNRAS.480.2578L}. 

\subsubsection{IR selected Compton thick AGNs}

A quantitative, complete assessment of the demography of highly obscured AGN with intermediate-to-high luminosity (the so called type 2 QSOs) is still lacking because building up complete samples of highly obscured, high-luminosity QSOs with homogeneous selection criteria is difficult and time-consuming, due to the large area that must be covered.
Complementary selection criteria combining far to near infrared to optical photometry, have been successful in pinpointing candidate obscured AGN. In particular, highly obscured AGN and QSOs can be selected by requiring extreme values of the 24 $\mu$m to optical flux ratio and red colors, which are demonstrated to be reliable proxies of high luminosity and high obscuration \citep{2008ApJ...672...94F}. For example, the SWIRE survey, covering with medium-deep MIPS and IRAC photometry about 50~deg$^2$ of the sky, provides a good opportunity to build up such a complete sample of highly obscured QSOs. PHEMTO observations of the brightest SWIRE sources with faint optical counterparts can easily probe their obscured nuclei. These sources are expected to host the most luminous and obscured AGN in the high-redshift Universe. The combination of X-ray and infrared information can be used to measure the number density of highly obscured QSOs. By joining this sample to those obtained from the deep fields, and from the serendipitous survey, we will be able to determine the evolution of the obscured AGN population, a step forward in completing the census of SMBH through Cosmic time.

\vspace{0.3 cm}



\subsection{Constraining explosion physics in supernovae and their remnants}


The optical light curve of supernovae (SNe) is powered by the decay of radionuclides synthesized at the nuclear burning front of SNe. Type Ia SNe, in particular, are the main source of Fe in the Universe, from the decay of their main product $^{56}$Ni to $^{56}$Co and $^{56}$Fe. The resulting nuclear lines and high-energy continuum emission (60-600~keV, which is produced by the Compton scattering of line photons) provide the most direct measurement of the nickel mass of SNe Ia, an essential constraint of the progenitor channel of SN Ia \citep{2012ApJ...760...54M}.

\begin{wrapfigure}{I}{0.5\textwidth}
\centering
\includegraphics[width=0.95\hsize]{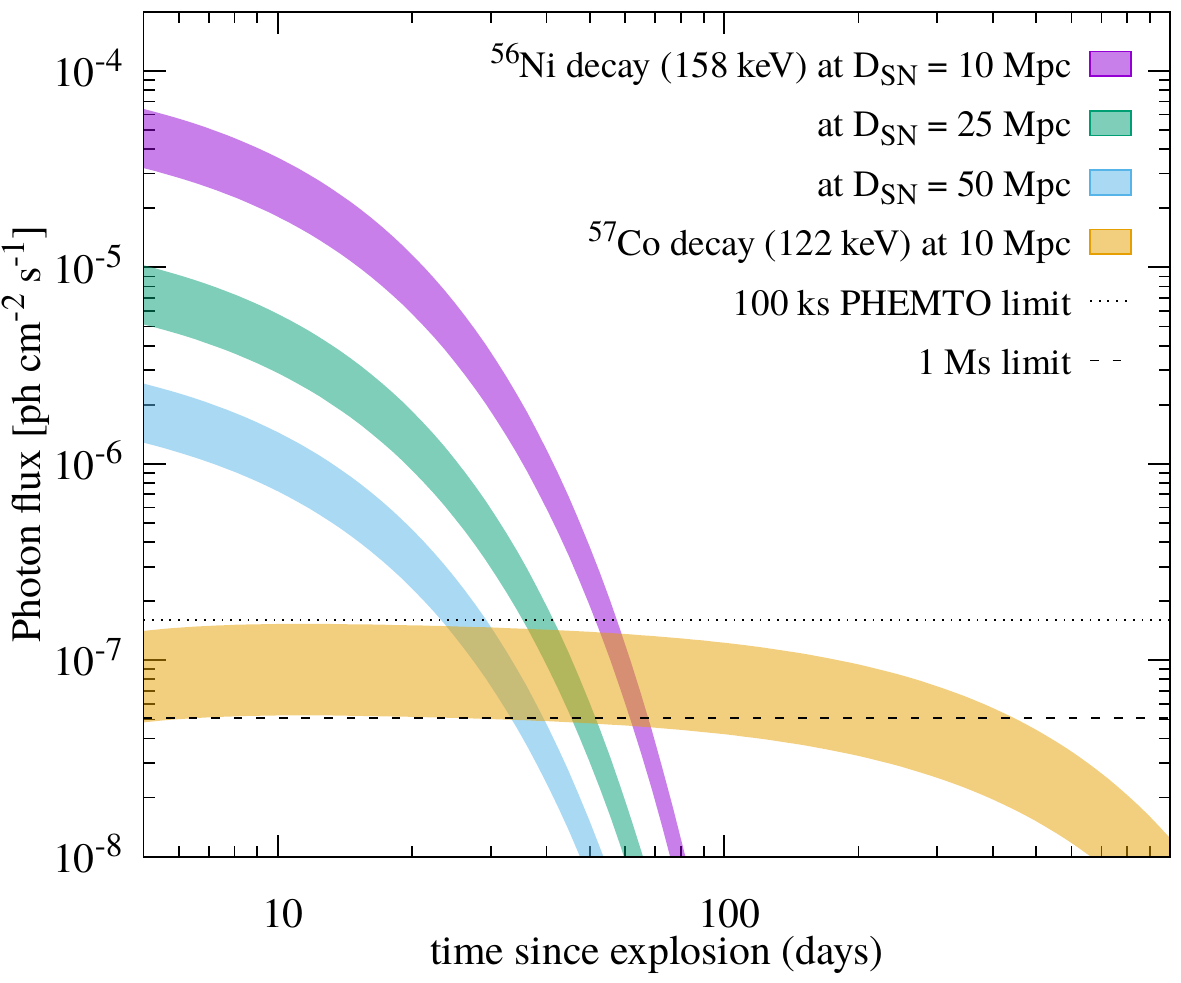}
\vspace{-1.0em}
\caption{\small Photon flux of decay lines of $^{56}$Ni (158~keV) from type Ia SNe at various distances. The spread corresponds to a range of Ni mass of 0.3-0.6~$M_{\odot}$. Shown in orange is the 122~keV line from the decay of $^{57}$Co for a range of $^{57}$Ni/$^{56}$Ni mass ratio from different explosion models \citep{2018ApJ...863..176M}. Estimated line sensitivity limits for 100 ks and 1~Ms exposures are shown in dotted and dashed line, respectively. No radiative transfer is assumed in these simulation, but the ejecta are transparent to hard X-rays ($>100$~keV) after 20 days \citep{1998MNRAS.295....1G}.}
\label{fig_SNIa}
\end{wrapfigure}

So far only the nearest type Ia SN2014J have been detected in $\gamma$-ray lines with INTEGRAL \citep{2014Sci...345.1162D}. As the 3D nature of the explosion and viewing angle affect the detection of radioactive decay $\gamma$-ray lines and measurement of Ni production, a larger sample of $\gamma$-ray-detected SN Ia is needed. For this, a combination of low background and high angular resolution and effective area is needed to dramatically increase the detection horizon: At an anticipated line sensitivity of $2\times 10^{-7}$~ph cm$^{-2}$ s$^{-1}$ for 100~ks, the 158~keV $^{56}$Ni $\rightarrow ^{56}$Co decay line can be observed routinely for SNe Ia within 50 Mpc (Fig.\,\ref{fig_SNIa}), or about one event per month. This population will include typical type Ia SNe but also sub- and super-luminous and so far exotic explosions \citep[SN1991bg-like or SN2002es-like, ][]{2017hsn..book..317T}, measuring for the first time directly the wide variety of Ni mass (0.1-1.4~$M_{\odot}$) predicted. 

Furthermore, the mass of another nickel isotope ($^{57}$Ni) can be observed from the 122~keV line of $^{57}$Co~$\rightarrow\,^{57}$Fe decay ($t_{1/2} = 272$~d). The ratio $^{57}$Ni/$^{56}$Ni is sensitive to the physical conditions (central density and metallicity) of the exploding white dwarf \citep{2018ApJ...861..143L}. For a SN Ia nearby ($<10$~Mpc, about one event per mission lifetime), we will have the possibility to obtain the first ever \textit{direct} measurement of $^{57}$Ni yield  (Fig.\,\ref{fig_SNIa}). Coupled with high spectral resolution from next generation X-ray instruments, which will measure the yields of $^{55}$Fe \citep{2015MNRAS.447.1484S} and the abundances of neutron-rich species (Cr, Mn) in supernova remnants \citep[SNRs, ][]{2019arXiv190405857W}, we will be able to fully characterize the physical conditions of the exploding WD and the mode of flame propagation (deflagration or detonation), and thus constrain directly both the progenitor channel of type Ia SNe and explosion physics.

\textbf{Probing stellar explosion with $^{44}$Ti\,:}
In the core-collapse (CC) SNe of massive stars, $^{44}$Ti is produced closest to the mass cut, the region separating material falling back on the newly-born neutron star from the ejecta. Its yield and the ratio  $^{44}$Ti/$^{56}$Ni are very sensitive to physical conditions \citep[peak temperature and density, ][]{2010ApJS..191...66M}, to properties of the progenitor \citep[rotation and metallicity, ][]{ChieffiLimongi}, and to the nature of the explosion  \citep[e.g. presence of bipolar jets, ][]{2016IJMPD..2530025N}.
In type Ia SNe titanium is produced in He shell or C/O cores but requires lower temperatures and densities ($T \lesssim 3\times10^9$~K and $\rho < 10^6$~g~cm$^{-3}$) for the material not to burn all the way to $^{56}$Ni \citep{2013ApJ...771...14H}. It is therefore more prominent for double-detonation models, in which a detonation of a thin He surface shell triggers core carbon/oxygen burning \citep{2010A&A...514A..53F}.

Since the decay of $^{44}$Ti to $^{44}$Sc (half-life of 59.2~yr) emits two lines at 68 and 78~keV, its yield can be measured in young supernova remnants up to a few hundred years old. The 3D distribution of this isotope can be revealed by resolving the emission both spatially (via imaging) and spectrally (via spectroscopy). Such measurements have been possible only for Cas A and SN1987A \citep{2014Natur.506..339G,2015Sci...348..670B}, but still suffers from large uncertainties. A jump in sensitivity by up to two orders of magnitude is needed to reap the benefits $^{44}$Ti can offer. It will be possible to compare the 3D dynamics of that element with that of iron, the decay product of $^{56}$Ni (obtained from high-spectral resolution X-ray observations, e.g. with Athena), to constrain many important aspects of the CC SN engine like the need for rapid rotation and/or jets \citep{2017ApJ...842...13W}. With this sensitivity jump the lines from $^{44}$Ti will be detectable for older Galactic CC SNRs such as Kes 75, 3C58, G21.5$-$0.9, and the Crab, for $^{44}$Ti yields of $(0.5-2) \times 10^{-4}~M_{\odot}$, thus offering new constraints on nucleosynthesis and explosion models \citep{ChieffiLimongi}. Since these remnants all contain pulsars or magnetars, the relative motion of titanium produced in the core of the star can be compared with that of the compact object to investigate the origin of neutron star kicks \citep{2017ApJ...842...13W}. Other Galactic remnants are from type Ia SNe (G1.9$-$0.3, Kepler, Tycho) and are close and young enough to allow resolving the distribution of $^{44}$Ti, accurately measuring its yield, and compare it to that of $^{56}$Ni. This will provide a discriminating diagnostic between progenitor channels that was so far out of reach due to limiting sensitivity \citep{2016MNRAS.458.3411T}. In particular, the ratio $^{44}$Ti/$^{56}$Ni is several orders of magnitude higher for the detonated He shell than from the core \citep{2018ApJ...854...52S}. Thus, mapping this ratio in type Ia SNRs would be the best direct test of the double-detonation model.

\vspace{0.3 cm}
\subsection{The role of magnetic field in cosmic accelerators and compact objects}


The fast shocks of young supernova remnants are efficient particle accelerators and the main sources of Galactic cosmic rays (CRs), as evidenced by the observation of synchrotron emission in the X-ray band. The same type of emission is seen from pulsar wind nebulae (PWNe), bubbles of relativistic particles (mostly pairs) inflated by pulsar winds and interacting with the surrounding medium, forming a relativistic shock which accelerates particles up to PeV energies. 
Although the basic concepts of diffusive shock acceleration are well understood, many important aspects remain unexplored, such as the maximum energy the particles can reach, the amplification of seed magnetic fields, the level of magnetic turbulence, and the magnetic configuration around neutron stars and in their winds . 
Because polarized X-rays are emitted by the highest energy electrons closest to the acceleration sites and are less affected by depolarization than in radio and optical, high angular resolution hard X-ray imaging and polarimetry is the ideal way to study the acceleration processes and measure the properties of the magnetic field in and around stellar remnants.

\subsubsection{The role of the magnetic field in SNRs and PWNe}

\textbf{ Particle acceleration and turbulence via the shape of the synchrotron cutoff }
\medskip

\begin{figure}
    \centering
    \includegraphics[width=18cm]{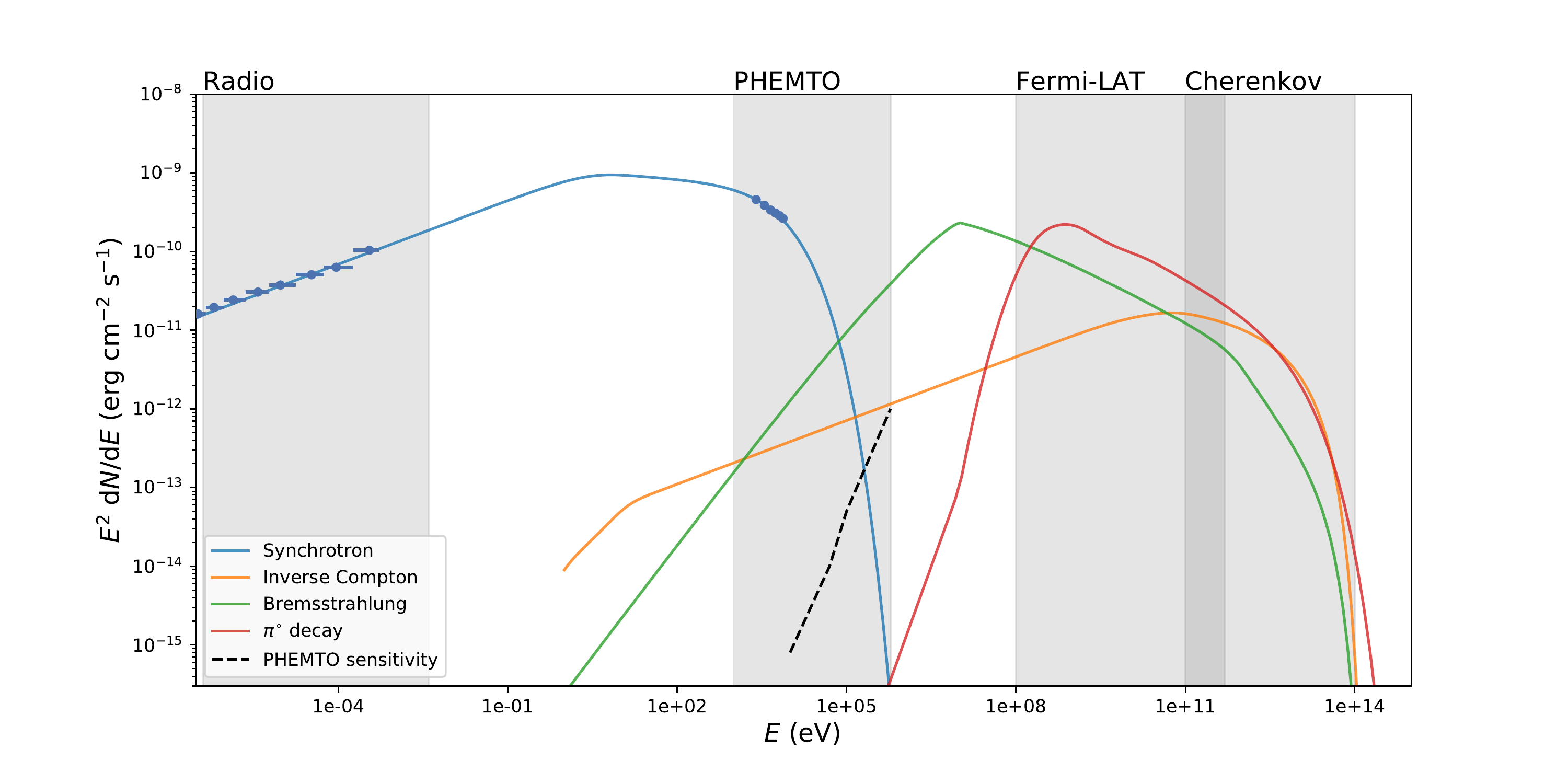}
    \vspace{-1.2cm}
    \caption{Spectral energy distribution showing the different radiation processes at play in SNRs. The radio and Chandra X-ray synchrotron data points from the Cassiopeia A SNR are used here to anchor the  electron population assuming a magnetic field of B=50 $\mu$G. Both the inverse Compton and the non-thermal Bremsstrahlung emission should be detectable by PHEMTO. 
}
    \label{fig:SNRSED}
\end{figure}

Observations of synchrotron X-rays prove that young supernova remnants accelerate electrons, perhaps to energies as high as 100 TeV. Yet the maximum energies these particles can reach and the details of the acceleration mechanism are still unclear.
For most young SNRs, the synchrotron X-ray emission is in the spectral cutoff region and its shape can provide important constraints on the acceleration mechanism and in particular on the diffusion coefficient $\alpha$ ($D \propto E^\alpha$). Fig.\,\ref{fig:SNRSED} shows the example of the young SNR Cassiopeia A where the synchrotron cutoff can be fully covered in the 1-600 keV energy band.

In a general form the electron population can be described as:

\vspace{-0.2cm} \begin{equation} \vspace{-0.2cm}
\mathrm{d} N /\mathrm{d}E  \propto E^{-s} \times \exp(- (E/E_{\rm cut})^{\beta} ).
\label{expcut} \end{equation}

where $s$ is the slope of the electron distribution and $\beta$ the curvature factor of the cutoff shape.
 In the case of loss-limited acceleration, the synchrotron cooling will balance acceleration and limit the maximum energy reached by electrons.
In this case, the cutoff curvature $\beta = \alpha +1 $ \citep{2007A&A...465..695Z} which leads to $\beta$=2 for the case of Bohm diffusion ($\alpha=1$).

We note that what we observe is the synchrotron radiation and not the particles properties directly.
The link between the curvature of the electron population ($\beta$) and the observed synchrotron spectral curvature depends on magnetic field configuration (uniform or turbulent) and needs to be constrained simultaneously by polarization measurements.
The polarization degree of synchrotron emission is in particular increasing with energy and peaks close to the electron cutoff energy \citep{2009MNRAS.399.1119B}.
The polarization degree of SNR synchrotron emission depends sensitively on the spatial scale of magnetic turbulence, as both the size of clumpy polarized structures and their polarization degree increase for steeper turbulence spectra. Such structures are variable on timescales of months to year \citep{2007Natur.449..576U,2009MNRAS.399.1119B} and hard X-ray monitoring of the flux and polarization in synchrotron-dominated SNRs (Cas A, RX J1713.7$-$3946) thus enable to constrain magnetic turbulence.

There are currently poor observational constraints on the curvature of the synchrotron cutoff because the leverage arm of ``soft" (0.3 to 10 keV) X-ray telescopes is too narrow. While the hard X-ray observations from NuSTAR have allowed to obtain a high-energy spectrum of the Cas A SNR (the brightest X-ray SNR), this spectrum was obtained from the entire SNR and individual filaments cannot be studied due to the limited sensitivity and angular resolution.
This large scale emission includes many physical processes that complicate the interpretation of the cutoff curvature, such as cooling, magnetic field evolution behind the shock, the time-integrated shock history, or a possible contribution from the reverse shock and contact discontinuity.
In order to probe the acceleration to the closest of the shock front and to allow a clear interpretation of the curvature measurement, extracting the spectrum from individual filaments at the shock is needed.
In young and bright historical SNRs (Cas A, Tycho, Kepler) the width of the filaments is of the order of 3-5$''$ and isolating filaments requires arcsecond-resolution extending above 10 keV.

\textbf{In short, a high-sensitivity, arcsecond-resolution telescope observing over a large X-ray band with polarization capabilities will allow us to constrain the cosmic ray diffusion and magnetic turbulence properties, key factors in the acceleration mechanism, in an unprecedented way.}

\medskip
\noindent \textbf{Magnetic field amplification structures: polarization in the precursor}
\smallskip

One of the main ingredient in the  diffusive shock acceleration process is the enhancement of the magnetic field turbulence upstream of the shock in the precursor  via the non-resonant cosmic ray return current instability (aka Bell’s instability). This enhanced turbulent magnetic field allows for the particles to be confined in the shock region, hence following multiple acceleration cycles and allowing them to reach high energies.
Despite major improvements in our understanding of this process using  MHD  and  PIC  simulations \citep[e.g.][and references therein]{2014ApJ...794...46C} and theoretical studies \citep{2017A&A...601A..64Z} there has been little observational constraints on the precursor properties.
Detecting directly the synchrotron radiation of accelerated particles in this synchrotron precursor would open a new window on the microphysics of the acceleration mechanism.
This emission is expected to be narrow with a typical scale of the order of a few 10$^{16}$~cm which translates into $\sim$1$''$ for a source at 1~kpc.
Despite its high angular resolution, the Chandra telescope has not been able to detect this precursor even in deep observations. One issue is to disentangle the precursor emission from that of a corrugated shock front. 
Polarization would be the essential tool to break this degeneracy as \textbf{the precursor is expected to have an increased polarization fraction with respect to that of the shock front} \citep{2017A&A...601A..64Z}.

In addition, the hard X-ray band (E$>$ 10 keV) is perfectly suited for this search as the size of the precursor scales with energy and its spectrum is expected to be harder than the shock front. Therefore a hard X-ray mission with polarization and arcsecond resolution will allow to tackle this long standing issue in a brand new way. 

\medskip
\noindent \textbf{Probing the inverse Compton emission in the hard X-ray band} 
\smallskip

The study of the inverse Compton $\gamma$-ray emission in SNRs can be used to probe the properties of the electron population and estimate the magnetic field strength by comparison with the X-ray synchrotron emission. However, this method suffers from several limitations. First, it is still a matter of debate in many SNRs whether the gamma-ray emission is dominated by emission from the electrons or hadrons (via the $\pi^{0}$ process, see Fig.~\ref{fig:SNRSED}) scrambling the link between the photon spectral properties and the electron properties.
In addition, due to the limited angular resolution of gamma-ray instruments, the spectral properties of electron population and the estimate of the magnetic field strength can only be obtained at the level of the entire SNR.

A high-angular resolution and high-sensitivity mission in hard X-rays is required to observe the IC and non-thermal Bremsstrahlung emission at 100 keV and above as shown in Fig.~\ref{fig:SNRSED}, therefore sampling the lowest part of the electron population and estimating the magnetic field strength by comparison with the radio synchrotron emission. While in some dense regions, the non-thermal Bremsstrahlung emission might dominate, we can study the IC component by isolating filaments evolving in a lower density environment with a high angular resolution mission.

\textbf{The 1-600 keV PHEMTO energy band has the unique capability to resolve and measure the properties of the highest and lowest energies of the electron population in a single observation. } 

\medskip
\noindent \textbf{Magnetic configuration in pulsar wind nebulae} 
\smallskip

\begin{wrapfigure}{I}{0.4\textwidth}
\centering
\includegraphics[width=0.99\hsize]{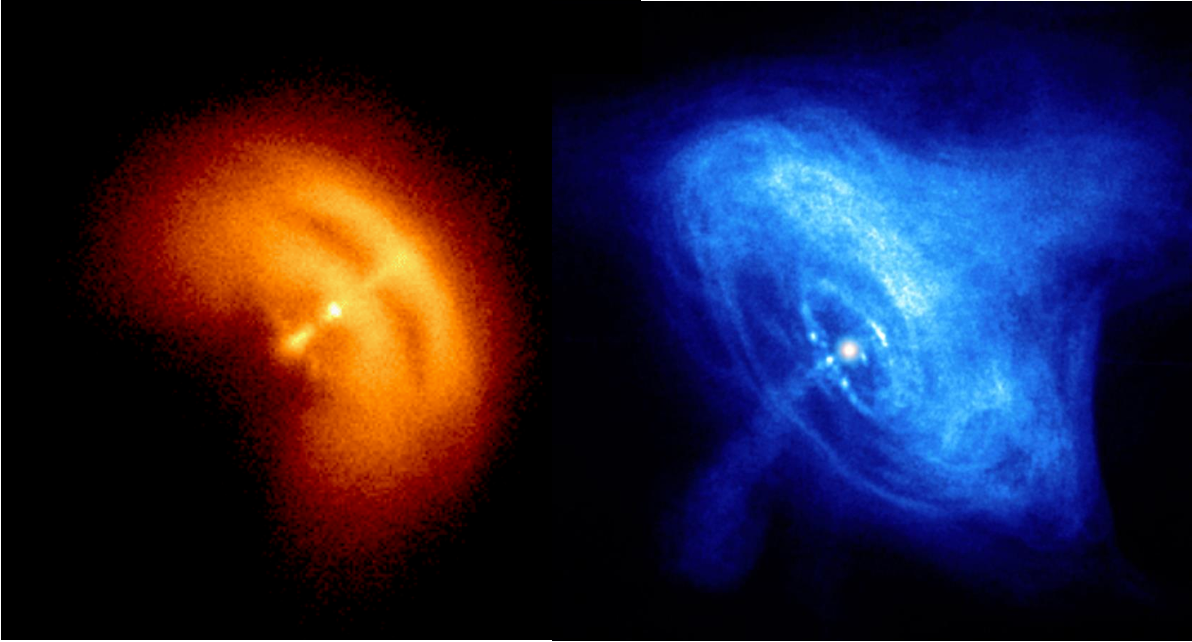}
\vspace{-1.0em}
\caption{\textit{Chandra} images of the Vela \citep[left,][]{2003ApJ...591.1157P} \citep[right,][]{2008ARA&A..46..127H} PWNe.}
\label{fig_PWN}
\end{wrapfigure}
X-ray images of of many young PWNe reveal axisymmetric features with a torus and jet, e.g. in the Vela and Crab nebulae (Fig.\,\ref{fig_PWN}). The torus can be explained by an anisotropic pulsar wind energy flux (higher in the equatorial plane) interacting with a toroidal magnetic field which is compressed and accelerates particles. However, to produce jets an additional magnetic component is necessary, either a poloidal or disordered (isotropic) field.

The synchrotron emissivity in PWNe is sensitive to both the magnetic field perpendicular to the fluid motion and the flow speed (via Doppler boosting), the latter making it difficult to constrain the field structure. Polarization thus provides a more direct mean of measuring the magnetic configuration, but the radio polarization already detected in PWNe is dominated by lower energy electrons from the outer regions. Radio polarization can discriminate ordered or disordered magnetic field for the overall nebula but not for the regions close to the termination shock, where the acceleration takes place. Only resolved X-ray polarization maps would allow to lift the degeneracy between Doppler boosting and magnetic field configuration (orientation, turbulence) closest to the acceleration site \citep{2017MNRAS.470.4066B}.

Due to the initial SN kick, older pulsars eventually escape the hot interior of their parent SNR and interact with the undisturbed interstellar medium, where the sound speed is much lower. Such supersonic PWNe develop a bowshock-tail or cometary structure \citep{2008AIPC..983..171K}. The wide variety of morphology observed is likely explained by the relative angles of the pulsar spin axis, kick direction, and the magnetic axis and wind energy flux. 3D MHD simulations have only recently attempted to assess the role of these parameters \citep{2019MNRAS.484.4760B,2019MNRAS.484.5755O}. High spatial resolution X-ray polarimetry is required to constrain the magnetic configuration and thus assess the important misalignment between spin axis and magnetic axis originating in the stellar collapse (see Sect.\,\ref{magnetar}).

\subsubsection{Magnetars as powerful Cosmic Accelerators}
\label{magnetar}
Magnetars are believed to be neutron stars (NS) whose magnetic field is several orders of magnitude larger than regular radio
pulsars (B$\sim$10$^{13-15}$G), and whose main energy source is the magnetic one.
Magnetars have been originally invoked to explain the phenomenology of Anomalous X-ray Pulsars (AXPs) and Soft-Gamma Repeaters (SGRs), that are relatively bright X-ray pulsars with slow periods (2-12 s) and high spin-down rates (10$^{-10}$-10$^{-12}$ s s$^{-1}$). These sources can be persistent or transient (undergoing outbursts lasting weeks to months), and some of them show short time variability emitting several $\sim$0.1 s bursts or event so-called giant flares where an extremely energetic short spike of gamma rays is followed by a ringing tail lasting for several hundreds of seconds.

Besides SGRs and AXPs, magnetars are nowadays believed to play a crucial role in a much wider range of astrophysical phenomena, from ultra-luminous X-ray 
sources (ULXs) and fast radio bursts (FRBs) to (cosmological) gamma-ray bursts (GRBs). The recent multi-messenger  observation  of  the  gravitational  wave  event  GW170817, has now testified that the coalescence of two neutron stars can, depending on the masses of the two objects and on the neutron star equation of state, result in the formation of a short-lived, rapidly rotating magnetar that could eventually collapse into a black hole once it slows down. 
Furthermore, evidence of strong magnetospheric activity from highly magnetized 
neutron stars are also observed in rotation-powered pulsars, and pulsating neutron stars in some X-ray binaries, showing that highly magnetic objects are indeed ubiquitous among different classes of sources.

Although magnetars are routinely observed in the soft X-rays with XMM-Newton and Chandra, the physics of their emission at high energies is still largely unknown. A dedicated spectro-polarimeter as PHEMTO can shed light on a number of hot unresolved topics.  

\medskip
- {\bf Physics of the magnetar high-energy emission.} Many magnetars exhibit high-energy hard tails that extend up to several hundreds of keV, either persistently or following transient events. These tails have been observed  by INTEGRAL, Suzaku and NuSTAR, but neither had the combination of sensitivity and spectral coverage required in order to study them in detail \citep[e.g.][]{gotz+06,enoto+17}, and hence the nature
of this emission remains unclear in terms of radiation and acceleration processes. SGRs and AXPs show different behaviours at high energies. While the spectrum of the former steepens, the latter show a spectral upturn. No spectral cutoff is observed, although its presence is inferred by COMPTEL upper limits. The basic picture for the high-energy magnetar emission involves the reprocessing of thermal photons emitted by the star surface through resonant Compton scattering (RCS) onto charges, moving in a twisted magnetosphere \citep[e.g.][]{nobili+08}. In addition, curvature radiation from ultra-relatistic particles can provide a substantial contribution. Crucially, the details are completely unknown: the distribution of the scattering particles in the velocity space is not understood as yet, nor is the geometry of the region where currents flow (the "\j-bundle"). These  affect dramatically the expected phase-dependent spectrum and polarization degree. A sensitive spectro-polarimeter like PHEMTO can nail down these  magnetospheric properties, reveal the magnetic field topology and locate the emission region(s) in the magnetosphere. It will allow us to map for the very first time the particle acceleration along closed field lines and to quantify its - highly non linear and not yet physically understood-  attenuation due to photon splitting and pair production in high B fields.  

\medskip
- {\bf System inclination and GW emission.} The application of models for the non-thermal magnetar emission to current data is also plagued by the poor knowledge of source inclination and viewing geometry. The expected polarization signature, however, depends significantly  on the geometry. Gamma-ray polarization measurements with PHEMTO will be crucial to reconstruct the system inclination (magnetic and spin axis with respect to the proper motion). PHEMTO will allow us to obtain information on the misalignment between spin and proper motion axis, which is still highly debated and is linked to the way in which the kick is imparted to a proto neutron star during its formation and to the duration of the physics of the acceleration phase. This is key to quantify the contribution of magnetars (including newly born ones) to GW emission, since orthogonal spin-velocity configurations will be efficient sources of gravitational radiation.  \textbf{Polarimetry observations in the  gamma-ray regime will complement the next generation X-ray polarimeters (as IXPE NASA and eXTP CAS) and the new generation of optical/infrared instruments designed for 30-m telescopes ultimately allowing for the very first time multi-wavelength  polarization studies of magnetars across the whole spectrum.} 

\medskip
- {\bf Make the first probe of the presence of newly born magnetars following GRB.} Some features observed in the afterglows of short Gamma-Ray Bursts (GRBs), as the so-called `X-ray plateau' could be explained by the presence of a magnetar after the coalescence of two compact objects which is believed to be at the origin of the short GRBs \citep[e.g.][]{stratta+18}. In addition this newly formed magnetar could be responsible for an isotropic transient X-ray emission due to the acceleration of particles in the rapidly rotating magnetosphere. In the 2050s ground based gravitational wave interferometers will routinely discover binary NS mergers up to redshift 1 and beyond \citep{ecm+11}: \textbf{follow-up observations with PHEMTO will contribute to elucidate the role of newborn magnetars in the formation of kilonovae and possibly provide a link between the young and evolved magnetar populations we observe in our Galaxy.}

\vspace{0.3 cm}
\subsection{Accretion and ejection physics}
Accretion is the most efficient way to produce energy and it powers a wide range of astrophysical objects at all ages of the Universe, ranging from distant gamma-ray bursts to close-by planet forming systems. It acts on all geometrical and temporal scales: from km-scale emission regions in the hearts of X-ray binaries (XRBs) to supermassive black holes in active galaxic nuclei and from (milli)second scales of GRBs and quasi-periodic oscillations in XRBs to millions of years lifetimes of protostars. The most violent and rapid manifestations of accretion phenomena are associated with black holes and neutron stars. Almost thirty years of multi-wavelength observations of accreting objects have shown that ejections (relativistic in the case of compact accretors) are seemingly always associated with specific phases of accretion through mechanisms that are yet to be understood and that are crucial in understanding all accreting phenomena. Studies of accretion in highly magnetized, young neutron star further offer also the possibility to directly probe the behavior of matter in extreme magnetic fields.

\subsubsection{Accretion and ejection physics in black holes on all mass scales}

\begin{wrapfigure}{I}{0.5\textwidth}
    \includegraphics[width=8cm]{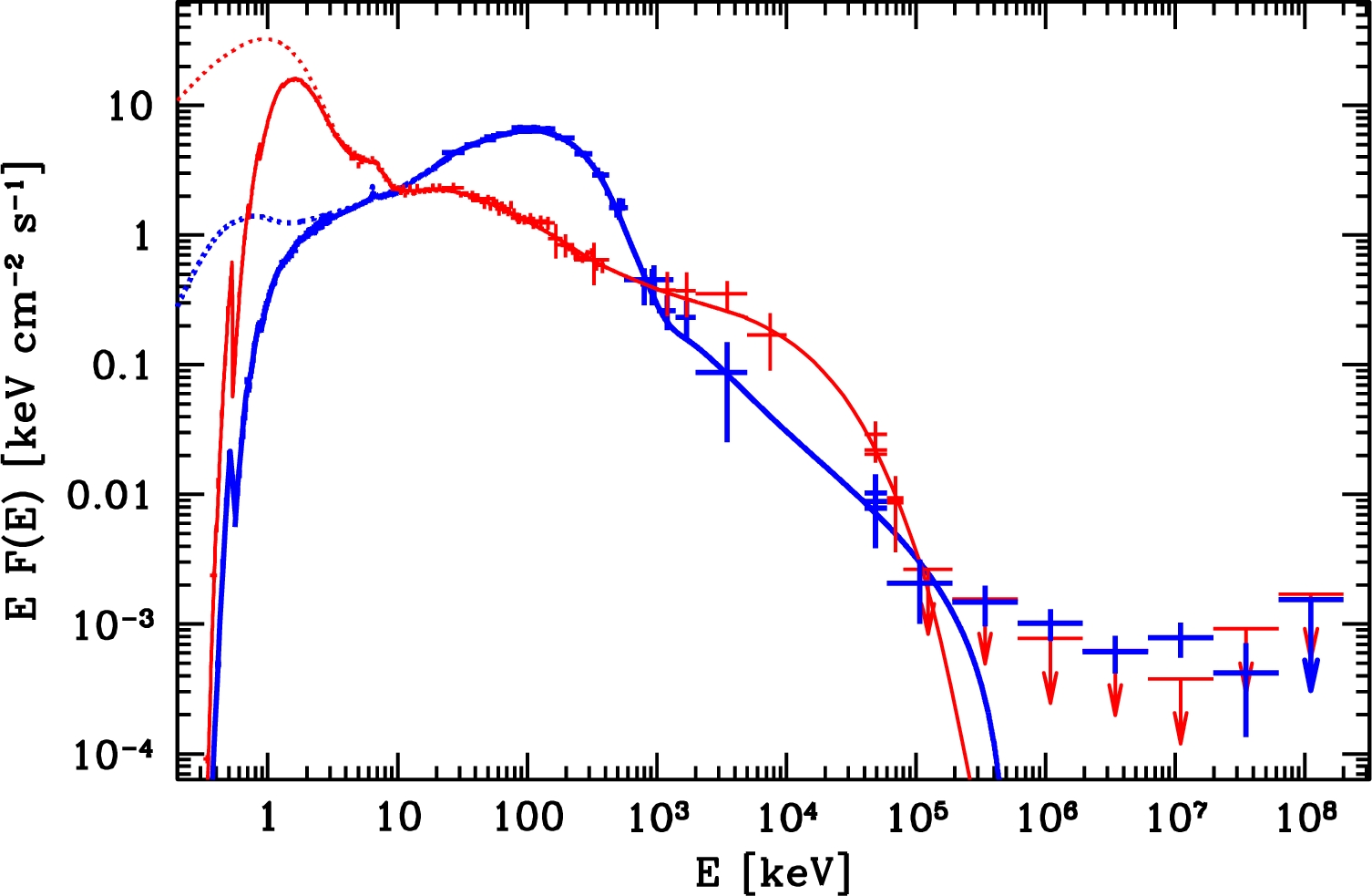}
    \caption{High-energy broad band spectra of the microquasar Cyg X-1 in 
    its main spectral states: the (low) hard state (blue) and the (high) soft state (red). While differences are obvious, especially in the 
    classical X-rays, the spectra show extensions at high energies (above 
    few hundred keV) whose origins are unknown \citep[from][]{zdziarski17}.}
    \label{fig:hespec}
\end{wrapfigure}

In AGN and XRBs, the bulk of the emission is released in  the X-ray and soft gamma-ray bands, i.e, typically in the range of $\sim$1\,keV to a few~MeV. The spectral window of PHEMTO is thus ideal to probe the physics of accretion. It has, however, become clear  over the past decade that the diagnostics enabled by spectroscopy alone are not sufficient to break the degeneracy between the various advanced theoretical models put forth to explain the broad band emission in accreting sources \citep{nowak11}. Polarimetry would provide a new, independent and complementary window onto the physics of accretion, providing access to various new parameters such as geometry of the system or strength of the magnetic field.


The currently accepted contributors to the accretion geometry in black holes and neutron stars with a low magnetic field are an accretion disk, 
a Comptonising medium (referred to as `corona', whose origin, geometry and global properties are largely unknown and highly model dependent), reflection features (Compton hump at a few tens of keV, highly broadened iron fluorescence feature at $\sim$6.4\,keV). They contribute to the overall emission in different proportion, resulting in different emission states, from hard states dominated by a power law component about $\sim$10\,keV to disk-dominated soft states (Fig.~\ref{fig:hespec}).
This already rather complex picture has been further complicated by the discovery of disc winds in disc dominated states, seen through absorption lines below $\sim10$\,keV \citep[e.g.][]{ueda98,neilsen09}, and hard tails (100 keV--10 MeV) in various states \citep[Fig. \ref{fig:hespec},][]{grove98,mcconnell00,laurent11,jourdain12,cangemi19}. The launching mechanism of the winds and the origin of the hard tails are highly debated.

For X-ray polarimetry, accretion into a deep gravitational well constitutes an unique science case: relativistic aberration and beaming, gravitational lensing, and gravito-magnetic frame-dragging  result in energy-dependent polarization signatures. \citet{connors80}   demonstrated that the polarization vector of light is transported in parallel along the photons’ null geodesic, while the degree of polarization is a scalar invariant. This results in strong predictions and thus direct tests for X-ray corona models; for example, uniform coronae should show polarization fractions around 2--10\% above 10\,keV.
PHEMTO will be ideal to test this for a handful (5--10) of XRBs and a few very bright disc/corona dominated AGNs where the coronal emission is expected to be polarized at typical values of $\sim$8\%, \citep{sunyaev85,schnittman10}.
For jet-dominated AGN, \citet{mcnamara09} have shown that X-ray polarimetry can distinguish between the three proposed scenarios to explain the origin of hard X-rays. The external Compton scenario in particular has a unique signature in both polarization degree and angle, e.g., it would show a much lower polarization fraction ($<$5\%) than the inverse-Compton and synchrotron 
self-Compton scenarios.
In blazars, the polarization degree might be very large, close 
to the theoretical limit given by uniform magnetic fields, so the observations of the polarization of synchrotron X-rays would allow to probe the structure of the magnetic field close to the base of
the jet, where the X-ray emission is produced.

The innovative potential of X-ray/soft gamma-ray polarimetry has been recently demonstrated with the  preliminary polarimetric studies allowed by the double layer detector of the IBIS telescope onboard INTEGRAL \citep{lebrun03,forot08}. Long-term monitoring observations of various sources such as the Crab pulsar/nebula, a couple of GRBs and one XRB (Cyg~X-1) have shown the presence of highly polarised emission above 400\,keV \citep{forot08, laurent11, goetz13, rodriguez15_cyg}, associated with synchrotron emission from the jet in GRBs and Cyg X-1 \citep[e.g.][]{goetz13,rodriguez15_cyg}, at least in jet dominated states, thus giving us a smoking gun for the origin of this emission. These detections are, however, burdened by large measurement uncertainties. INTEGRAL as a polarimeter has a low sensitivity while covering only the energy range above a few hundred keV, only enough to foreshadow the potential of polarimetic methods.

For example, both leptonic and hadronic models for blazars predict X-ray/gamma-ray polarization signatures \citep[e.g.,][]{Zhang_2013} but the level of polarization above $\gtrsim$10\,keV is substantially higher for hadronic processes. Combined with multi-wavelength polarization observations, PHEMTO will thus distinguish leptonic from hadronic scenarios in AGN relativistic jets. High-energy neutrino detection is another way of confirming hadronic acceleration in relativistic jets \citep[e.g.,][]{2018Sci...361.1378I}: relativistic protons interacting with synchrotron photons produce both neutrinos and secondary particles whose synchrotron emission leaves an imprint in the 40\,keV--40\,MeV energy range \citep{10.1093/mnras/stu2364}. Features of this process, the Bethe-Heitler pair production, are efficient ways of constraining hadronic acceleration in blazar relativistic outflows; by detecting them PHEMTO will play a central role 
in proving the origin of high-energy cosmic rays.

\begin{wrapfigure}{I}{0.5\textwidth}
	\includegraphics[height=14cm]{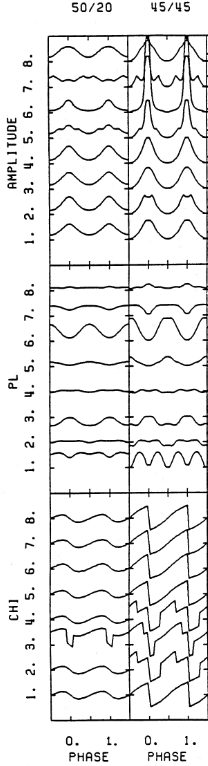} \includegraphics[height=14cm]{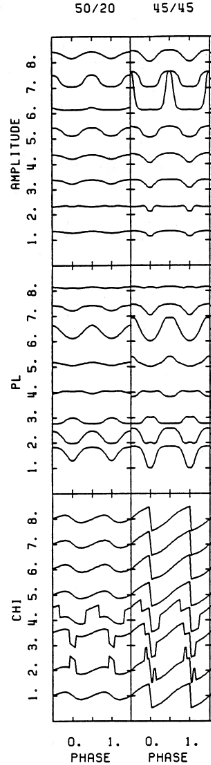}	
	\caption{The \emph{left panel} shows from top to bottom the pulse phase amplitude, the polarization degree, and the polarization angle for an X-ray pulsar emitting a pencil beam from two antipodal accretion columns. In each panel, the lines from top to bottom refer to different energies (from 84.7 to 1.6 keV, the third one from the top corresponds to the cyclotron energy at 38.4 keV). The two columns of this panel correspond to two different orientations of the spin and magnetic axes from the line of sight. The \emph{right panel} is for a fan beam.  \citep[from][]{Meszaros1988}.}
	\label{fig:hmxb-polarization}
\end{wrapfigure}

In summary, \textbf{thanks to PHEMTO's broad energy coverage and spectroscopic and polarimetic abilities we will be able to constrain the drivers of the accretion/ejection processes.} Through temporal variability, broad band spectroscopy, and polarimetry we will be able to characterize the accretion flow and study causal relations and interplay between the various media. \textbf{PHEMTO particularly has the unique ability to reveal the geometry and physical origin of the individual accretion/ejection components (disc, jets, corona, disc winds) and their mutual influence; polarimetry will enable to break the degeneracy between competing models.} \textbf{PHEMTO will further study the impact of the magnetic field onto accretion/ejection processes, with a particular focus on its evolution and its relationships to the different accretion states and thus accretion geometries.}



\subsubsection{Accretion in highly magnetized neutron stars}

An important subclass of X-ray binaries comprises a high mass star and a neutron star as accretor: these are 
know as \emph{neutron-star high-mass X-ray binaries} (NS-HMXB). Since the neutron star is relatively young, it is still endowed in a strong magnetic field of the order of 10$^{12}$\,G which might imprint characteristic signatures in the hard X-ray spectrum, known as cyclotron line scattering features \citep[see][for a review]{Staubert2019}.   

These features appear as absorption lines with a large width of a few keV at an energy between 10 and 100 keV on the exponentially decaying part of the spectrum. They constitute the only direct measurement of the neutron star magnetic field in its immediate vicinity. Despite this phenomenon being know since 1976, a lively debate is still present on the location of the line-forming region and on the actual geometry of the accretion flow producing the X-ray emission. For instance, it is not clear if 
cyclotron scattering features are formed at the boundary of the magnetically confined accretion column, or on the neutron star surface \citep{Becker2012,Poutanen2013}; and which is
the relative contribution and localization of the accretion columns \citep[see e.g.,][for a recent attempt to characterize the accretion geometry using energy-dependent pulse profiles]{Iwakiri2019}. 

It has been shown long ago \citep{Meszaros1988} that polarization properties of X-ray pulsar are energy-dependent and change dramatically when the photon energy crosses the cyclotron energy. This is due to the birefringence of plasma embedded in a magnetic field and has the potential to constrain the geometry of the accretion stream. \citet{Meszaros1988} has also shown that the polarization pattern is different if the radiation is emitted at the top of the accretion column (pencil beam) or from the sides of the columns (fan beam). For such a study, it is thus essential to measure polarization at different spin phases of the neutron star: in a pencil beam configuration, the polarization angle changes from positive to negative in a discontinuous manner \emph{at pulse maximum}, whereas in a fan-beam, this happens at pulse minimum (see Fig.~\ref{fig:hmxb-polarization}).

Observing such pattern would yield a definitive prove of our current understanding for the changes of pulse profiles that are observed at different luminosity in several sources \citep[see e.g.][for an early interpretation of this phenomelogy]{Parmar1989}. Even more strikingly, polarization must be different for reflection models and models with lines originating in the accretion stream itself, 
but a quantitative estimation has not been published, yet. Analysis of polarization data will benefit from the development of spectral physical models for the accretion column \citep[e.g.][]{Becker2007,Farinelli2016} combined with the geometrical constraints derived from
the study of pulse profiles \citep{Iwakiri2019}. 

The current generation of high energy polarimeters does not cover the cyclotron lines energy range (IXPE is limited to 8 keV, for instance). {\bf With its large spectral coverage and polarimetric capabilities, PHEMTO has the unique potential to explore the cyclotron scattering features forming region in great details and thus understand the behavior of matter in extremely high magnetic fields.}
\vspace{0.3 cm}
\subsection{Hard X-ray emission of galaxy clusters}


The presence of relativistic particles and magnetic fields in the intra-cluster medium (ICM) of galaxy clusters has been established by a number of observations at radio frequencies \citep{Feretti12,vanWeeren19} and poses important physical questions \citep{Brunetti14}. The extended ($\sim$Mpc), diffuse, low surface brightness, steep-spectrum synchrotron sources known as radio halos are produced by relativistic electrons spiraling around $\sim \mu$G magnetic fields. Such halos have so far been detected in $\sim$50 clusters at $z<0.4$, and in a few other clusters at higher redshift  \citep{Yuan15}. All these clusters are characterized by high mass, high X-ray luminosity and temperature, and they show indications of a merger process as probed by X-ray morphology \citep{Buote01,Cassano10}, X-ray temperature maps \citep{Govoni04} and presence of optical substructures \citep{Girardi11}. A causal connection between the hot and relativistic plasma properties is then suggested by the similarities of the radio morphology of the halos with the X-ray structure of the ICM \citep{Govoni01a}, and the total radio halo power $P_{1.4GHz}$ at 1.4 GHz correlates with the cluster total X-ray luminosity $L_{X}$  \citep{Giovannini09,Kale13}. The radio power also correlates with the total cluster mass \citep{Cassano13}.

The synchrotron emission is a combined product of both the particle and the magnetic field density, therefore the latter cannot be globally constrained by radio observations alone. However, the same population of relativistic electrons that produce the radio emission will also produce hard X-ray emission by inverse Compton (IC) scattering photons from the ubiquitous Cosmic Microwave Background (CMB). For a power-law energy distribution, the ratio of IC to synchrotron flux gives a direct and unbiased measurement of the average magnetic field strength in the ICM. The search for non-thermal IC emission in clusters began with the first X-ray sensitive satellites, although the extended $\sim$keV photons from clusters were soon recognized to be of thermal origin \citep{Mitchell76}. However, clusters showing a diffuse and extended radio emission must also present IC emission at some level. Unfortunately, non-thermal emission is hard to detect, due to thermal photons which are simply too numerous below 10 keV, while IC emission should dominate and produce excess emission at higher energies, where the bremsstrahlung continuum falls off exponentially. In addition, the presence of multi-temperature structures, naturally occurring in merging galaxy clusters, could cause a false IC detection and needs to be modeled accurately; this problem is particularly relevant when using non-imaging and high background instruments, like the ones commonly used for the search of non-thermal emission in the hard X-ray band. The first IC searches with \textit{HEAO-1} resulted only in upper limits, and thus lower limits on the average magnetic field of $B \gtrsim 0.1 \, \mu$G \citep{Rephaeli88}. The next generation of hard X-ray satellites instead - \textit{RXTE} and \textit{Beppo-SAX} - claimed a few detections ($10^{-11} - 10^{-12}\, \mathrm{erg}\, \mathrm{s^{-1}}\, \mathrm{cm}^{-2}$), although mostly of marginal significance and controversial \citep{Rephaeli08}. More recent observatories however (\textit{Suzaku} and \textit{Swift}) did not confirm IC emission at similar levels \citep{Ajello10,Wik12,Ota14}. \textit{NuSTAR} \citep{Harrison13}, the first satellite with imaging capabilities in the hard X-ray band (3-79 keV) is clearly making pioneering observations of few iconic clusters such as the Bullet cluster \citep{Wik14} or the Coma cluster \citep{Gastaldello15} clearly revealing that the average magnetic field in radio halos clusters is typically closer to $\sim 1 \mu$G than the 0.2 $\mu$G implied by past detections. 

Only a satellite with the capabilities of PHEMTO will be available not only to detect but to {\bf map} with spatially resolved spectroscopy the IC emission therefore being able to reconstruct the spatial distribution of the relativistic electrons and the magnetic field, in a way complementary to the RM constraints provided by SKA. {\bf In this way we will be able to reconstruct the physical properties of the non-thermal particles of galaxy clusters to the same level of accuracy likely achieved for the thermal ICM in 2050.}

The dissipation of the total kinetic energy involved in cluster mergers occurs through turbulence, thought to be the driver of the formation of radio halos \citep{Brunetti14}, and shocks. Shock fronts are unique observational tools to study the physical processes in the ICM. In particular, they can be used to determine the velocity and kinematics of the merger and to study the conditions and transport processes in the ICM, including thermal conduction and electron-ion equilibration. The temperature profile across a shock front can, indeed, disentangle the heating mechanisms of electrons and the microphysical processes at work. An instantaneous rise in temperature at the front and a flat profile behind indicates electrons are directly heated at the shock itself, in contrast to expectations from Coulomb interaction arguments where electrons largely ignore the shock and are initially only adiabatically compressed before gradually equilibrating with the ions. Those measurements are really challenging due to the fact that high shock temperatures ($kT \gtrsim 10$ keV) are poorly constrained by the low effective area above 5 keV of Chandra and XMM and the relatively poor angular resolution of NuSTAR. Recent studies could not resolve with the needed precision the degeneracy between electron shock heating models nor give stringent upper limits on the effective thermal conductivity in post-shock regions \citep{Markevitch06,Russell12,Markevitch03}. \textbf{The effective area at high energies and 1" spatial resolution will allow these crucial measurements of the ICM physics to be made routinely on a large sample of clusters.}

By 2050 ICM calorimeter data taken by XRISM and Athena will be fully exploited. It is an easy guess to predict that many new discoveries will be made and that the complex nature of the ICM will be even more revealed, making the definitive transition from an ideal gas to a complex low beta plasma description, as for example departure from a Maxwellian for the emitting electrons as commonly observed in the Sun's corona and heliosphere and Earth's magnetosphere. \textbf{A broad band description taking the ICM spectra to high energies and temperatures (above the Athena bandpass) will be clearly crucial for further progress in our understanding.}

\newpage

\section{Scientific Requirements}
\begin{wrapfigure}{I}{0.45\textwidth}
    \includegraphics[width=\hsize]{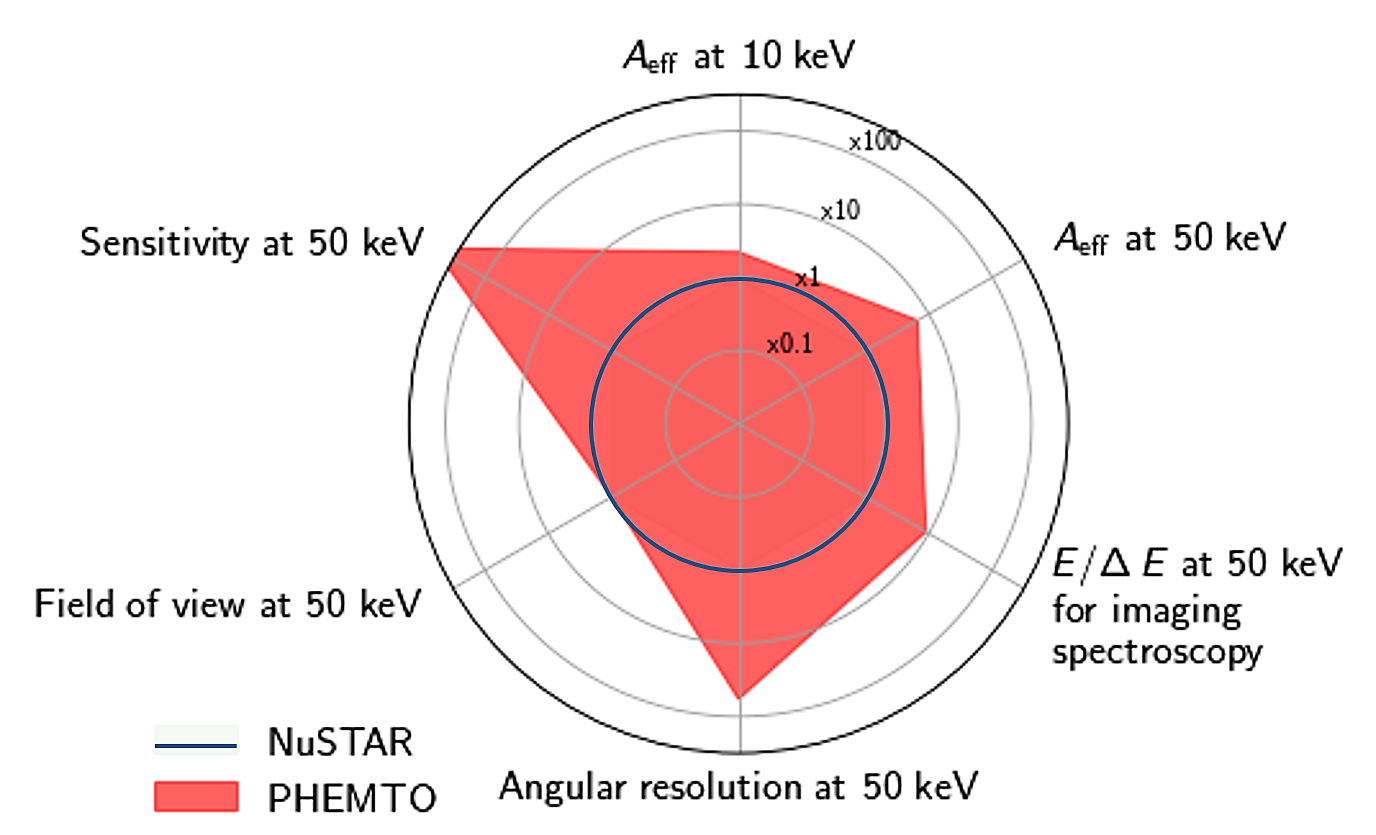}
    \caption{comparison of PHEMTO with the \textit{NuSTAR} mission (see \citet{Harrison13} for NuSTAR performances assessment).}
    \label{fig_radar}
\end{wrapfigure}

To address these main scientific objectives, the PHEMTO scientific requirements are given in the table below. We show also in Figure \ref{fig_radar} the comparison of the main PHEMTO characteristics with the NuSTAR mission. With these characteristics, the PHEMTO observatory will also address many astrophysical questions in Solar System physics, Stellar astrophysics, studies of the Interstellar Medium, ... and, without any doubts, many serendipitous objectives and questions which may be opened in the next decades.

\vspace{0.1 cm}

\renewcommand{\arraystretch}{1.5}
\begin{table}[h]
\centering
\caption {PHEMTO scientific requirements}
\vspace{0.1 cm}
\begin{tabular}{|c|l|c|}
\hline
\textbf{Requirement} & \textbf{Parameter} & \textbf{Value} \\ \hline
SR-1 & Energy band                     & 1 - 600 keV  \\ \hline
SR-2 & On-axis continuum sensitivity   & $10^{-16}$ erg~cm$^{-2}$~s$^{-1}$ @ 10 keV \\
 &  & $10^{-14}$ erg~cm$^{-2}$~s$^{-1}$ @ 100 keV \\
 &  & $3~ 10^{-13}$ erg~cm$^{-2}$~s$^{-1}$ @ 600 keV \\
\hline
SR-3 & Minimum Detectable Polarisation & 1 \%     \\
\hline
SR-4 & Field of view                   & 6 arcmin   \\
\hline
SR-5 & Angular resolution (H.E.W.)     & $\approx$ 1$''$   \\
\hline
SR-6 & On-axis effective area          & 2000 cm$^2$ @ 10 keV  \\ 
&           & 500 cm$^2$ @ 600 keV  \\ \hline
SR-7 & Spectral resolution             & E/$\Delta$E = 100 @ 100 keV \\
\hline
SR-8 & Detectors background            & $\leq 10^{-4} \rm \, cts \, s^{-1} \, cm^{-2} \, keV^{-1}$ \\
\hline
SR-9 & Absolute timing accuracy        & $50~\mu$s \\
\hline
SR-10 & Time resolution                & $50~\mu$s \\
\hline
SR-11 & Mission duration               & 5 years \\
\hline
\end{tabular}
\label{table_req}
\end{table}

\newpage

\section{Mission concept}

\subsection{Overview of all proposed payload elements}

The PHEMTO observatory is based upon two focusing devices, mirrors at low energy and a Laue lens at higher energy, focusing X–rays onto two identical focal plane detector systems.

\subsection{Summary of mirror payload key resources and characteristics} 

The envisioned technical implementation for the PHEMTO observatory is to have two independent telescopes looking simultaneously at the same sky. One more efficient at low energy, based on focusing mirrors working in reflection conditions, and the other one at higher energies, based on the Laue lens technique. On the low energy branch, the gain in the maximum energy that can be focused will be achieved by having a very long focal length, up to 100 m. This focal length will be obtained most probably in a formation-flying configuration.  

\vspace{0.2 cm}

\textbf{Mirror optics }

We are considering two main mirror technologies for the PHEMTO mission : Silicon Pore Optics and Glass optics. Silicon pore optics (SPO) are made of silicon mirror plates diced from standard silicon wafers. The plates are ribbed on one side (called the back side). The front side is used for the reflection of X-rays. Then the ribbed plates are bonded to form a stack. Doing so, a stack forms a set of concentric conical surfaces. In order to approximate a Wolter-I system, two stacks are placed one after the other in such a manner that two SPO stacks form a cone-cone system. SPO optics have the advantage to have a very good reflective area / mass ratio while keeping excellent performances on the angular resolution ; this is why they have been chosen for the mirror of the ATHENA mission of ESA, with a 5 arcsec angular resolution (HEW) requirements. 

The glass optics technology (GOp) has been successfully used for the hard X-ray mirror assemblies of the NASA NuSTAR mission. The mirror substrates are thin sheets of flexible glass which start out as flat sheets. The glass is heated in an oven and slumped over precisely polished cylindrical quartz mandrels to achieve the right curvature. The slumped mirror segments are then coated with a multi-layer process. The optics are built from the inside out, shell upon shell, spaced apart by graphite spacers and held together by epoxy. They presently allow to reach energies up to $\sim$ 80 keV, with an angular resolution of $\sim$ 1 arcmin (HEW). 

\vspace{0.2 cm}

\textbf{Laue lens}

Laue lenses exploit the crystal diffraction in transmission configuration (Laue geometry). A Laue lens is made of a large number of crystal tiles in transmission configuration, that are disposed in such a way that they concentrate the incident radiation onto a common focal spot. Laue lens work well at high energy up to 600 keV and more. A detailed description of the Laue lens principles and lens development status is given in  \citet{Frontera13, Virgilli18} and in the ASTENA (Advanced Surveyor of Transient Events and Nuclear Astrophysics) mission description (Frontera et al. 2019 in preparation). Two white papers (one on transient events led by Cristiano Guidorzi and another on Nuclear Astrophysics led by Filippo Frontera) based on the ASTENA concept have been submitted to ESA for the Voyage 2050 long term plan.

\subsection{Summary of detector payload key resources and characteristics} 

\subsubsection{Description of the measurement technique}

The PHEMTO focal plane (FPA) is designed to detect single photons focused by the mirrors in the 1 -- 600 keV energy range. To do so, FPA is equipped with two superimposed semiconductor based imaging spectrometers, LED (Low Energy Detector - potentially made in Silicium) and HED (High Energy Detector - potentially made in CdTe). Both are embedded into an active and passive shield system, AS (Anticoincidence System). LED and HED will measure the interaction position, energy deposit, and arrival time of each incoming X-ray. The LED detector will operate in the low energy range from 1 up to 40 keV, whereas the HED will cover the 8 to 600 keV range. The FPA is also a hard X-ray Compton Polarimeter, by combining LED/HED coincidence data. A collimator will be placed on top of the FPA to stop all photons coming out of the mirror field of view. AS will allow detecting charge particles of the space environment, avoiding a significant loss of sensitivity due to induced background on LED and HED. The whole payload system will be calibrated in-flight by means of radioactive sources mounted on a calibration wheel. A protective enclosure surrounds the detector assemblies to provide environmental protection and to block stray light.  

\subsubsection{Detector Payload conceptual design and key characteristics}

\noindent \textbf{Low Energy Detector (LED) description and characteristics}

LED operates from 1 up to 40 keV range. It may consist of a $2048 \times 2048$ matrix, with a $100 ~\mu$m pixel size. This leads to a $22 \times 22$~ cm$^2$ Silicon monolithic imager with a double sided process. Each pixel will correspond to 0.2$''$ in the sky. The LED will be a matrix permanently read out in a rolling shutter mode, and equipped with a Fast Trigger (LET). The LET is a self‐trigger system to get accurate time of arrival of particles. It is also used to perform fast coincidence analysis with the HED for Compton event detection and anticoincidence with the AS events for background rejection.

\vspace{0.2 cm}

\noindent \textbf{High Energy Detector (HED) description and characteristics}

HED operates between 8 and 600 keV. It fully covers the mirrors high-energy capabilities. It may consist of a $2048 \times 2048$ CdTe matrix of $22 \times 22$ cm$^2$ size. The pixel size should be also $100 \mu$m.

\vspace{0.2 cm}

\noindent \textbf{Anti-Coincidence Detector description and characteristics}

The anticoincidence system provides accurate tagged time trigger (100 ns) for particles crossing the detectors or matter close to the detectors, which generate internal background. It is fully opaque to photons up to 200 keV, outside of the FoV and self-absorbs secondary fluorescence photons (passive graded shield). The active part of the system will have more than 99\% efficiency for passing-through minimum ionizing particles.

\vspace{0.2 cm}

\subsection{State of the art and foreseen technological developments} 

\subsubsection{Low Energy Detectors}

The PHEMTO mission could take benefit from the latest technologies of silicon detectors, in particular large arrays of active pixel sensors developed for the Wide Field Imager of ATHENA \citep{Meidinger17}. However, one important required improvement would be the timing resolution in order to implement Compton events analysis for polarimetry studies and an effective anticoincidence system to reduce the instrumental background. Some concepts of fast detectors based on DEPFET technology are under study with small demonstrators \citep{Treberspurg19} and could result in real concepts for astronomy in the next decade. The introduction of a self-triggering front-end electronics architecture (LET) in addition or instead of the nominal rolling shutter readout mode is also a promising field of research.

\subsubsection{High Energy Detectors}

The requirements of both small pixel size ($100 \mu m$) and large arrays ($22 cm^2 \times 22 cm^2$) are very challenging and require developments in rupture. Hard X-ray imaging detectors with $50 \mu m$ pitch are already available \citep{Ruat19,Poikela14} but with poor spectral resolution not suitable for space astronomy. The promising developments of fine-pitch hard X-ray spectroscopic imagers imply CdTe double-sided strip detectors ($150 \mu m$ pitch, \citet{Furukawa19}) and CdTe pixel detectors ($250 \mu m$, \citet{Wilson15}). A dedicated R\&D program will be of high importance to reach the science goals of the PHEMTO mission and includes several development axis that could be shared between European research labs and industrial in 2020-2040.
These developments are in synergy with industrial motivations for medical applications (cancer tomography…), synchrotron and free electron laser facilities or homeland security (luggage scanning). Joint decadal R\&D program roadmaps are also possible in this sector.

\subsubsection{Anticoincidence System}

The Anticoincidence system will also benefit from new technologies, either, for instance, in the making of 3D printed graded shields which could efficiently surround the whole FPA with optimal mass and no leaks, or in using Artificial Intelligence inboard to efficiently suppress background events.

\subsubsection{Mirror Optics}

SPO have been chosen for the mirror of the ESA/ATHENA mission, with a 5 arcsec angular resolution (HEW) requirements. It is reasonable to estimate that this new technology, never flown for now, will be maturing enough in the next decades to allow reaching a 1 arcsec angular resolution performance below $\sim$ 10 keV. 
On the other hand, on the NuSTAR misison, GOp with a Pt/C coating presently allows to reach energies up to $\sim$ 80 keV, with an angular resolution of $\sim$ 1 arcmin (HEW). As for SPO, it is anticipated that the GOp and coating technologies can progress sufficiently in two decades to reach the angular resolution required for PHEMTO up to a few hundred keV. 

\subsubsection{Laue Lens}

The PHEMTO mission will take benefit of the development of Laue lens made with bent Silicon and Germanium crystals.  Bent crystals are currently produced and the technology development is described in \citet{Virgilli18}. The development of a lens prototype is the goal of a recently approved  project TRILL (Technological Readiness level Increase for Laue Lenses) devoted to increase its Technology Readiness Level (TRL). In a next future, we can reasonably expect to achieve a 10 arcsec angular resolution (HEW) at a 100  meters focal distance. We anticipate that the technology can progress in the next 20 years to reach our scientific requirements.

\newpage

\vspace{3cm}

\noindent\textbf{Core team}:  P.~Laurent$^{1}$, F.~Acero$^{1}$, V. Beckmann$^{2}$, S. Brandt$^{3}$, F. Cangemi$^{1}$, M. Civitani$^{4}$, M. Clavel$^{5}$, A. Coleiro$^{6}$, R. Curado$^{7}$, P. Ferrando$^{1}$, C. Ferrigno$^{8}$, F. Frontera$^{9}$, F. Gastaldello$^{10}$, D. Götz$^{1}$, C. Gouiff\`es$^{1}$, V. Grinberg$^{11}$, L. Hanlon$^{12}$, D. Hartmann$^{13}$, P. Maggi$^{14}$, F. Marin$^{14}$, A. Meuris$^{1}$, T. Okajima$^{15}$, G. Pareschi$^{4}$, G.W. Pratt$^{1}$, N. Rea$^{16}$, J. Rodriguez$^{1}$, M. Rossetti$^{10}$, D. Spiga$^{4}$, E. Virgilli$^{9}$, S. Zane$^{17}$ 

\vspace{3cm}

\noindent\textbf{Institutes}:\\

{\footnotesize
\noindent $^1$ CEA/IRFU/DAp, CEA, CNRS, Universit\'e Paris-Saclay, 91191 Gif-sur-Yvette, France\\
$^2$ CNRS/IN2P3, 3 rue Michel Ange, 75016 Paris, France\\
$^3$ DTU, Elektrovej, 2800 Lyngby, Danemark \\
$^4$ INAF, Osservatorio Astronomico di Brera, via E. Bianchi 46, 23807 Merate, Italy\\
$^5$ IPAG, Universit\'e Grenoble Alpes, 38058 Grenoble Cedex 9, France \\
$^6$ APC, CNRS/Universit\'e de Paris, CEA, Observatoire de Paris, 75205 Paris, France\\
$^7$ Physics Department, University of Coimbra, Coimbra, Portugal\\
$^8$ ISDC, University of Geneva, chemin d’Ecogia 16, 1290 Versoix, Switzerland\\
$^9$ Department of Physics and Earth Science, University of Ferrara, Via Saragat 1, I-44122 Ferrara, Italy\\
$^{10}$ INAF, IASF-Milano, via Alfonso Corti 12, 20133 Milano, Italy \\
$^{11}$ Institute for Astronomy and Astrophysics, Universit\"at T\"ubingen, Sand 1, 72076 T\"ubingen, Germany \\
$^{12}$ University College Dublin, School of Physics, Science Centre Belfield, Dublin 4, Ireland \\
$^{13}$ Clemson University, Department of Physics \& Astronomy, Clemson, S.C. 29634-0978, USA \\
$^{14}$ Universit\'e de Strasbourg, CNRS, Observatoire astronomique de Strasbourg, UMR 7550, 67000 Strasbourg, France \\
$^{15}$ NASA/GSFC, Code 662, Greenbelt , MD 20771, USA \\
$^{16}$ Institute of Space Sciences, CSIC-IEEC, 08193 Barcelona, Spain \\
$^{17}$ Mullard Space Science Laboratory, University College London, United Kingdom \\
}

\newpage

\bibliographystyle{aa}
\bibliography{references}
\end{document}